# VERITAS Observations of the Galactic Center Ridge

A. W. Smith for the VERITAS Collaboration*

*University of Maryland, College Park, College Park, MD, 20742 USA*

*NASA Goddard Space Flight Center, Greenbelt, MD, 20771, USA*

*veritas.sao.arizona.edu

Due to its extraordinarily high concentration of known relativistic particle accelerators such as pulsar wind nebula, supernova remnants, dense molecular cloud regions, and the supermassive black hole (Sgr A*); the center of the Milky Way galaxy has long been an ideal target for high energy (HE, 0.1-100 GeV) and very high energy ( VHE, 50 GeV-50 TeV) gamma-ray emission. Indeed, detections of Sgr A* and other nearby regions of gamma-ray emission have been reported by EGRET and Fermi-LAT in the HE band, as well as CANGAROO, Whipple, HESS, VERITAS, and MAGIC in the VHE band (see [1] for a summary). Here we report on the results of extended observations of the region with VERITAS between 2010-2014. Due to the visibility of the source for VERITAS in the Northern Hemisphere, these observations provide the most sensitive probe of gamma-ray emission above 2 TeV in one of the most complicated and interesting regions of our home galaxy.

## 1. THE GALACTIC CENTER RIDGE

The Galactic Center region is perhaps one of the most interesting local regions for study in the very high energy (>100 GeV) gamma-ray band. This is primarily due to its high concentration of star forming regions, pulsar wind nebulae, supernova remnants, and of course the central accelerator Sgr A*; all of which are known sources of VHE gamma rays. The region has been studied extensively in both the TeV band, as well as the GeV band with Fermi-LAT [refs]. Due to its high density of possible gamma-ray sources, the confirmed number of individual sources (point or extended) is relatively low (<5), with a very large proportion of gamma-ray emission in the region coming from either unresolved point sources, or a diffuse, extended component. Observations of this region with HESS [refs] reveal a distinct band of emission stretching along the central region of the plane; this emission seemingly correlated with dense molecular cloud regions. As this diffuse component is assumed to be generated from cosmic ray interactions with the molecular clouds, the study of this region in the TeV band also allows for a characterization of the cosmic ray flux near the Galactic Center.

In addition to the conventional gamma-ray sources in the Galactic Center, this region is also believed to be the closest, densest concentration of particle dark matter in our local universe. If the lightest supersymmetric particle ($\chi$, or the neutralino) is indeed the correct explanation for particle dark matter, the Galactic Center should present a flux of GeV-TeV gamma rays due to $\chi\chi$ self-annihilations. While this flux is model dependent, some dark matter models place it within the sensitivity of current detectors, while many more will be probed by the upcoming CTA observatory. Regardless, the principal limiting factor in the use of observations of the Galactic

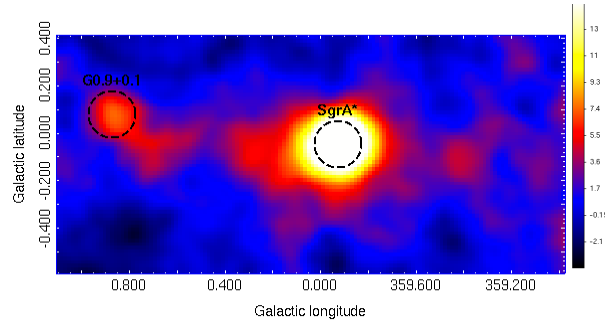

Figure 1: The VERITAS 2D significance map of the Galactic Center Ridge, smoothed with the $0.12^0$ PSF of the instrument for these observations. The central source is saturated due to high significance. Also shown in black dashed circles is the VERITAS PSF for the observations.

Center for constraining dark matter is the poorly understood nature of the conventional gamma-ray sources in the region. Therefore, a better understanding of the gamma-ray source population in the Galactic Center can also provide insight into indirect searches for particle dark matter.

In this proceeding we present preliminary results from the VERITAS observations of the Galactic Center region in the >2 TeV regime. These results confirm many of the previous HESS results in the >300 GeV region, but due to the higher energy range of the VERITAS observations, also provide a unique window on the highest energy gamma-ray emission in the Galactic Center region.

## 2. VERITAS OBSERVATIONS

The Very Energetic Radiation Imaging Telescope Array System (VERITAS), located in Southern Arizona (USA) is an array of four 12-meter imaging atmospheric Cherenkov telescopes (IACTs) providing excellent angular resolution and sensitivity to cosmic TeV sources. In normal operations (i.e. high elevation observations), VERITAS is sensitive in the energy range of 80

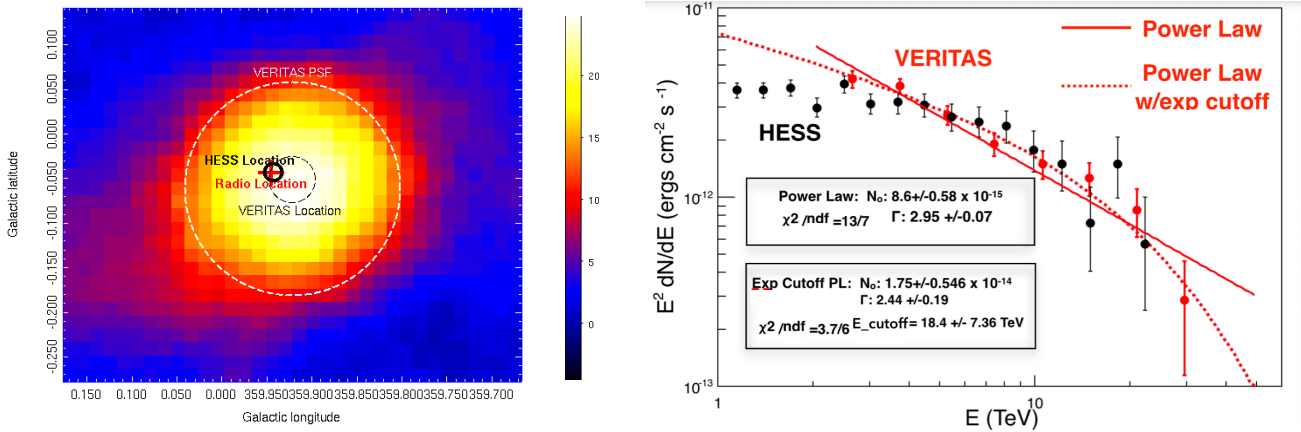

Figure 2: The smoothed VERITAS 2D significance map of the Sgr A* (left). The right panel shows the VERITAS (red) and HESS (black) spectral points derived from the Sgr A* observations along with the fits described in the text. Note, the fits were made to VERITAS points only.

GeV to 50 TeV and has the capability to detect a 1% Crab Nebula flux in approximately 25 hours of observations. VERITAS has an energy resolution of 15% at 1 TeV and a typical angular resolution of <0.1⁰ .

Between 2010-2014, VERITAS accrued ~85 hours of quality selected, livetime observations of the Sgr A* region. Due to the Northern Hemisphere location of VERITAS, the Sgr A* region never transits above 30⁰ elevation. This large zenith angle to the source results in a higher energy threshold (>2 TeV) for VERITAS observations. Normally, such observations would result in very poor angular resolution for ground based gamma-ray telescopes. However, using a specialized analysis technique (see [2][3][4]) in which the displacement between the center of gravity of a parameterized Hillas ellipse and the location of the shower position within the camera plane is utilized. This displacement or "DISP" method results in enhanced angular resolution at large zenith angle observations where small parallactic displacements between shower images would normally degrade angular resolution significantly (see [1] for a description).

## 3. RESULTS

### 3.1. Sgr A*

In the 85 hours of observations taken between 2010-2014, VERITAS strongly detected gamma-ray emission above 2 TeV from Sgr A*. A total of 735 excess gamma-ray events were detected by VERITAS, resulting in a detection significance of >25σ). The resulting 2 dimensional significance map, as well as the derived energy spectrum are shown in Figure 2.

The VERITAS position of the Sgr A* is in agreement with both the radio and HESS source locations [5]. The differential energy spectrum is fit by both a simple power law of the form $N_0$ x $(E/10$ $TeV^{-\Gamma})$·as well as a power law with an exponential cutoff of the form  $N_0$ x $e^{(-E/E_{cutoff})}$ x$(E/10$ TeV$)^{-\Gamma}$ We find that the exponential cutoff power law model provides a better fit (reduced $\chi2$ of 0.6 vs 1.8). The cutoff energy of 18 (+/-7.4) TeV is higher than that reported in [6].    Since VERITAS will continue observing the Sgr A* region at energies above 2 TeV, VERITAS will be able to more accurately constrain the cutoff energy of the Sgr A* spectrum a crucial parameter to physical models of emission from the source in both hadronic/leptonic as well as plerionic/accretion scenarios.

### 3.2. G0.9+0.1

The composite SNR ( X-ray plerionic core + radio shell) has previously been detected by the HESS collaboration [7] as a relatively weak source (2% of the Crab Nebula flux above 300 GeV). The current VERITAS observations also detect G0.9+0.1 as a significant TeV source above 2 TeV with a total of 134 excess counts, corresponding to a >7σ source detection. The VERITAS source position for G0.9+0.1 is consistent with both the plerionic core as well as the HESS location. The derived energy spectrum of the source is well fit by pure power law with an index of Γ= 2.3+/-0.1, with no indications of an energy cutoff up to >25 TeV.

### 3.3 Ridge Emission

In [5], the HESS collaboration presented the residual maps (i.e. after subtracting known point sources within the field of view) of the >300 GeV emission from the Galactic Plane. These residual maps revealed a complicated network of diffuse gamma-ray emission within the central $3^0$ of the Galactic Plane. When plotted along with the CO emission contours (see [5]), the HESS emission appears correlated with dense molecular cloud regions (bright in CO line emission). However, given the complicated nature of the region, this measurement was unable to rule out the possibility of a significant contribution to the TeV flux coming from unresolved point sources.

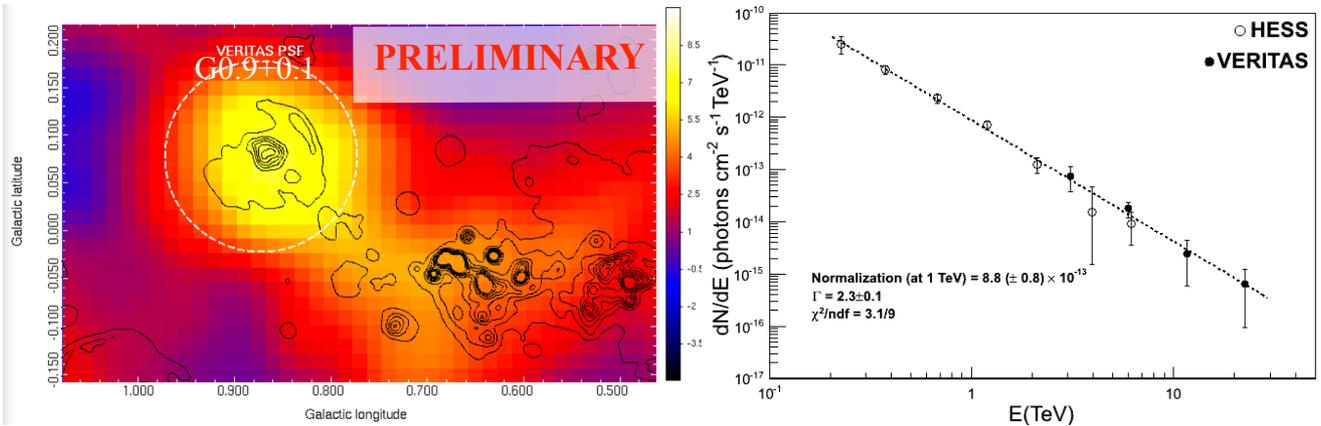

Figure 3: The smoothed VERITAS 2D significance map of the G0.9+0.1 region along with ARCO 20cm radio contours (left). Also shown (right) is the VERITAS differential energy spectrum of G0.9+0.1 along with the HESS spectral points from [7].

In Figure 4 is shown the VERITAS >2 TeV residual significance maps of the inner Galactic Center region after the point source emission from Sgr A* and G0.9+0.1 has been removed. It is clear from these maps that a diffuse component of TeV emission is present above 2 TeV both directly adjacent to Sgr A*, as well as extending ~0.80 to the left of Sgr A* along the Galactic plane.

Figure 4 also shows the HESS 300, 325, and 350 excess event contours (green), the ARCO 20cm radio emission contours (black), and the point sources from the 3FGL catalog (blue). As can be these maps that a diffuse component of TeV emission is present above 2 TeV both directly adjacent to, and extending ~0.8⁰ in Galactic longitude to the left of Sgr A*. seen there are direct correlations with both the HESS >300 GeV emission, as well as co-locations of 3FGL sources [8]. While there appears to be a good correspondence between the 20cm emission at the location of the Sgr B2 star forming region, the correlation between radio emission and > 2 TeV emission directly adjacent to the location of Sgr A* is less obvious. An upcoming publication will provide further examination of these residual maps.

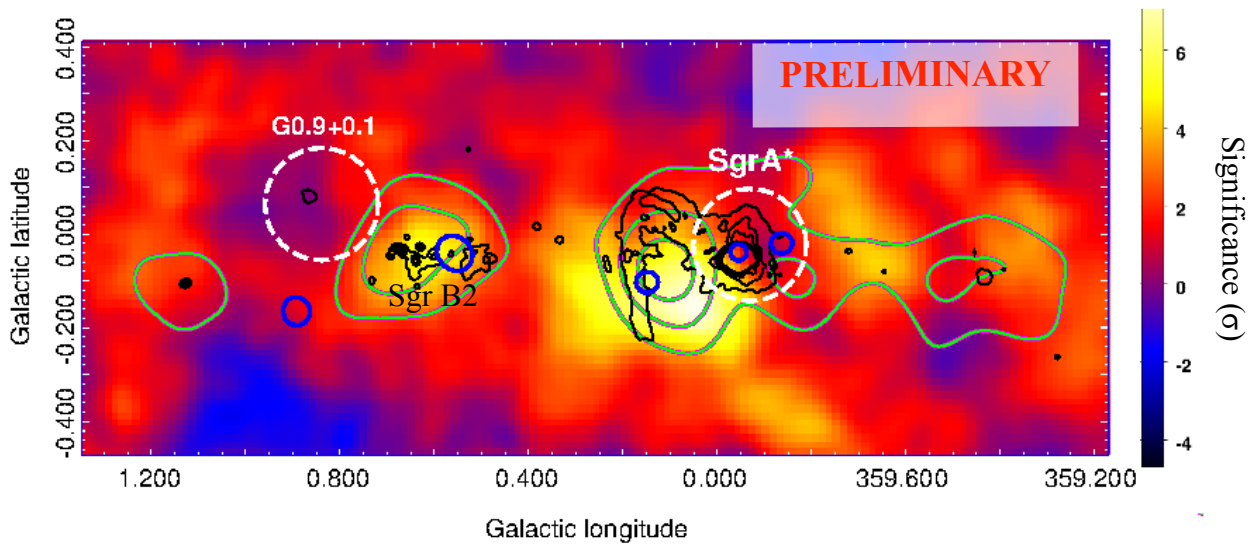

Figure 4: The smoothed VERITAS 2D residual significance map after emission from both Sgr A* and G0.9+0.1 have been removed (white dashed circles). Blue circles represent Fermi-LAT point sources from the 3FGL catalog, 20cm radio contours (ARCO) are shown in black, and the HESS >300 GeV excess event contours are shown in green.